\documentclass[journal,letterpaper,twocolumn]{IEEEtran}
\usepackage{amsmath,amssymb,graphicx,psfrag,cite,epsfig}
\usepackage[normalem]{ulem}
\usepackage[usenames,dvipsnames]{color} 
\usepackage{balance}
\usepackage{mathrsfs}   
\usepackage{paralist}
\usepackage{bbm}

\usepackage{xcolor}
\usepackage[linesnumbered,ruled,vlined]{algorithm2e}

\SetCommentSty{mycommfont}

\SetKwInput{KwInput}{Input}                
\SetKwInput{KwOutput}{Output}              

\graphicspath{{figure/}}

\usepackage{color}

  \renewcommand{\sout}[1]{}  

\newtheorem{thm}{Theorem}
\newtheorem{lem}{Lemma}
\newtheorem{cor}{Corollary}
\title{Parity Check Codes for Second Order Diversity}

\author{Aaqib A. Patel,~\IEEEmembership{Member,~IEEE,} Abdul Mateen Ahmed,~\IEEEmembership{Student Member,~IEEE,}
 and Mohammed Zafar Ali Khan,~\IEEEmembership{Senior~Member,~IEEE,} 
\thanks{Aaqib A. Patel, Abdul Mateen Ahmed  and Mohammed Zafar Ali Khan are with the Department of Electrical Engineering, Indian Institute of Technology, Hyderabad. e-mail: \{aaqib,ee15resch02001,zafar\}@iith.ac.in.}%
}%

\begin{document}
\maketitle

\begin{abstract}
 Block codes are typically not used for fading channels as soft decision decoding is computationally intensive and hard decision decoding results in performance loss. In this paper we propose a diversity preserving hard decision decoding scheme for parity check codes (PCC) over Rayleigh fading channels. The proposed flip decoding scheme has linear complexity in the block length. Theoretical analysis and simulation results verify the correctness of the proposed detection scheme.
\end{abstract}


\section{Introduction}
\label{sec:intro}
Brute-force maximum-likelihood decoding (MLD) of a linear $(n, k)$ block code requires calculation  of  $2^k$ metrics. This method becomes too complex to be applied for large $k$ and so more effective methods are required.  As linear block codes have a trellis structure \cite{1}, the Viterbi algorithm can used to reduce the number of computations. Yet, the  branch complexity of the states becomes large as $k$ increases.   Block code maximum likelihood decoding has been investigated by many coding theorists; as   detailed in \cite{2}. Most initial works trade-off optimal performance to reduce decoding complexity. In generalized minimum distance (GMD) decoding \cite{3}, an algebraic decoder is used to generate a list of codeword candidates. This list is determined by the reliability measures of the symbols within each received block. For each candidate, a test is then performed, which has a sufficient condition for optimality. The most likely candidate is chosen as the decoded codeword. Chase improved on this idea with an algorithm that searches for error patterns corresponding to the $t$ least reliable bits, where $t$ is the number of errors that the code can correct \cite{4}.
For reliability less than a predefined threshold for a given position, the error performance depends on the threshold \cite{5}. The number of calculations depends on the choice of threshold and the signal-to-noise ratio (SNR). These algorithms suffer in performance  as the code length increases. \cite{6} gives an  MLD algorithm   that does not place limits  on the search space at the beginning. But at each iteration, a new condition is tested for optimality until convergence to a specific solution. This optimal algorithm improves the computational complexities of \cite{4,5} for short codes, but increases exponentially with the dimensions of the code.

Another technique \cite{7} is to decode the syndrome and then use this information to improve the hard-decision decoding.
This schemes orders the information bits based on their credibility. Using this approach, different search schemes based on binary tree and graph were presented \cite{8}. However, the methods presented in \cite{7,8} need $n-k$ to be relatively small.  For very high-rate codes, a method to reducing the search space  was presented in \cite{9}.
Other methods take advantage of code structure to reduce the overall complexity of Trellis Decoding \cite{10,11,12,13}. Nevertheless, the trellis complexity still increases exponentially with the code length \cite{14}. To maintain complexity,  suboptimal multistage decoding on  trellis  has been proposed \cite{15,16}.
%

Ordered statistics-based decoding (OSD) is a soft-decision decoder for linear block codes. OSD provides a near maximum-likelihood (ML) performance  \cite{19,20}.  But suffers from  large computational complexity. Algorithms that reduce the  complexity of OSD are  referred to in the literature \cite{21,22,23,24,25}.

In this letter we propose a diversity preserving linear complexity hard-decision decoding for parity check codes. The proposed decoding has linear complexity in the block length, $n$. As PCC have a rate of $n-1/n$, this allows use of high rate codes that achieve second order diversity.  The contributions of this paper are
\begin{itemize}
	\item use of parity check codes for error correction over fading channels.
    \item a diversity preserving hard-decision decoding scheme for parity check codes.
    \item the proposed flip decoder (FD) has linear complexity.
	\item theoretical analysis that shows that the proposed decoder achieves second order diversity over Rayleigh fading Channels.
	\item simulation results that verify the theoretical analysis.
\end{itemize}

%
The rest of the paper is organized as follows. Relevant previous work is summarized in Sec.~\ref{sec:prev}. The proposed flip decoder is presented Sec. \ref{sec:flip}, where the diversity order of the decoder is also derived. Simulation results are presented in Secs.~\ref{sec:sim} and Sec.~\ref{sec:cncl} concludes the paper.
\section{Previous Work}\label{sec:prev}
\subsection{System Model}
Let $C$ be a $(n,k)$ binary linear code with generator
matrix $G$ and minimum Hamming distance $d_{\min}$. The code, $C$, is used for error
control over the Rayleigh fading channel. Let $\bar{c} = (c_1, c_2, \cdots,c_n)$ be a codeword in $C$.
The $n$ bits of the codeword are sent on a Rayleigh fading channel with ideal interleaving.  
 For BPSK modulation, the codeword $\bar{c}$ is mapped to a bipolar sequence $\bar{x} = (x_1,x_2,\cdots,x_n)$ where $x_i \in \{-1,1\}, \forall i$.
The  received signal can be represented as
\begin{equation}\label{eq1a}
  r_i = h_i x_i + w_i,1 \le i \le n
\end{equation}
 where $w_i$'s are statistically independent complex Gaussian random variables
with zero mean and variance $N_0$ and $h_i \in \mathbb{R}$'s are independent, identically distributed fading coefficients.
\subsection{Decoding and diversity}
If a hard-decision decoding is
performed, then the probability of error is upper bounded as \cite{Pro}
\begin{equation}\label{eq2a}
  P(c) \le \sum_{q=t+1}^{n}\binom{n}{q}\left(\frac{1-\beta}{2}\right)^{q}\left(\frac{1+\beta}{2}\right)^{n-q}
\end{equation}
where $\beta=\sqrt{\frac{\bar{\gamma}_c}{\bar{\gamma}_c+1}}$, $t$ is the number of errors that can be corrected by the code, $\bar{\gamma}_c$ is the average signal to noise ratio (SNR) per code bit and the instantaneous SNR per code bit is given by $\gamma_{l}=h_l^2 \bar{\gamma}_c$.  Note that the average SNR per bit is defined as $\bar{\gamma}_b=\bar{\gamma}_c/ R$ where $R$ is the rate of the code.  We have
\begin{lem}\label{lem1}
Hard-Decision decoding has a diversity order of $t+1=\left\lfloor{\frac{d_{\min}-1}{2}}\right\rfloor+1$
\end{lem}
\begin{IEEEproof}
Rearranging \eqref{eq2a}, we have
\begin{equation*}\label{eq2}
  P(c) \le \left(\frac{1-\beta}{2}\right)^{t+1}\sum_{q=0}^{n-t+1}\binom{n}{q}\left(\frac{1-\beta}{2}\right)^{q}\left(\frac{1+\beta}{2}\right)^{n-t+1-q}.
\end{equation*}
As the average SNR tends to a large value, $\lim_{\bar{\gamma}_c \rightarrow \infty} \left(\frac{1-\beta}{2}\right) = \frac{1}{4\bar{\gamma}_c}$ and $\lim_{\bar{\gamma}_c \rightarrow \infty } \left(\frac{1+\beta}{2}\right)^{q} = 1$.
Using the definition of diversity order, $D$, \cite{Tse} and substituting from above, we have
\begin{eqnarray}\label{eq3}
  D &=& \lim_{\bar{\gamma}_c  \rightarrow \infty} \frac{-\log [P(c)]}{\log[\bar{\gamma}_c ]} \nonumber \\
   &=& \lim_{\bar{\gamma}_c  \rightarrow \infty} \frac{(t+1)\log[4\bar{\gamma}_c]+\log\left[1+\frac{1}{4\bar{\gamma}_c }+\cdots\right]}{\log[\bar{\gamma}_c ]}\nonumber\\&=&t+1.
\end{eqnarray}
\end{IEEEproof}
While hard-decision decoding has lower complexity, it results in a loss of diversity order. This loss of diversity order has a significant impact on performance over fading channels.

Using soft decision decoding,  the probability of error is upper bounded as \cite{Pro}
\begin{equation}\label{eq3a}
  P(c) < \left(2^k-1\right)\left[\frac{4}{2+\bar{\gamma}_c}\frac{1+\bar{\gamma}_c}{2+\bar{\gamma}_c}\right]^{d_{\min}}.
\end{equation}
We have
\begin{lem}\label{lem2}
Soft-decision decoding has a diversity order of $d_{min}$.
\end{lem}
\begin{IEEEproof}
Using the definition of diversity order, $D$, \cite{Tse} and substituting from \eqref{eq3a}, we have
\begin{eqnarray}\label{eq4}
  D &=& \lim_{\bar{\gamma}_c \rightarrow \infty} \frac{-\log [P(c)]}{\log[\bar{\gamma}]} \nonumber \\
   &=& \lim_{\bar{\gamma}_c \rightarrow \infty} \frac{d_{\min}\log[2+\bar{\gamma}_c]+\log\left[\frac{1+\bar{\gamma}_c}{2+\bar{\gamma}_c}\right]+\log\{4(M-1)\}}{\log[\bar{\gamma}_c]}
   \nonumber\\&=&d_{\min}.
\end{eqnarray}
\end{IEEEproof}
 However, soft decision decoding requires comparison of all $2^k$ codewords, which becomes very costly as $k$ increases.
\subsection{Parity Check Codes}
 The  parity check codes are the most well known error-detecting code. In this linear code,  the codeword consists of  the  $k$-bit data word with an extra bit, called the parity bit, to get an $n= k + 1$-bit codeword. The parity bit  makes the total number of 1s in the code word even or odd. Without loss of generality, in this paper the parity is chosen to be even. The minimum Hamming distance for this code is $d_{\min} =2$. This means that the code  can detect a single error and it cannot correct any error.  The generator matrix and the parity check matrix are given by
 \begin{equation}\label{eq1}
    G=\left(
        \begin{array}{cc}
          I_k & 1_k
        \end{array}
      \right),
      H^T= 1_n
 \end{equation}
where $I_k$ is an identity matrix of size $k$ and $1_n$ is a column vector $n$ 1's. Lemmas \ref{lem1} and \ref{lem2}, specialize for this case, so that
\begin{cor}\label{cor1}
The  parity check codes have a diversity order of one and two  with hard-decision decoding with soft-decision decoding respectively.
\end{cor}
\begin{IEEEproof}
Substituting $d_{\min}=2$ in Lemmas \ref{lem1} and \ref{lem2} we have the desired result.
\end{IEEEproof}
 Note that Corollary \ref{cor1} implies that the diversity order of two can be achieved with PCC with soft decision decoding. However, as the code length increases the decoding complexity increases exponentially. In the next section we propose a hard-decision decoder for PCC that achieves full diversity of $d_{\min}=2$.
\label{sec:system}
\section{The proposed hard-decision decoder for PCC} \label{sec:flip}
Since the CSI is known at the receiver, if there is a parity mismatch, the most likely bit in error would be the bit with the smallest CSI, $h$. Based on this intuition we have the proposed decoder in Algorithm \ref{alg1}.
\begin{algorithm}[h]
\DontPrintSemicolon
  \KwInput{
  \begin{itemize}
    \item[\checkmark] $r$: received word of length $n$
    \item[\checkmark] $h$: absolute value of CSI
  \end{itemize}
  }
  \KwOutput{
  \begin{itemize}
    \item[\checkmark] $c$: the corrected codeword
  \end{itemize}
  }
  $[h_{sort},Idx] = sort(h)$;\tcp*{sort the CSI}
  \tcp*{check the parity }
  \If{$mod(sum(r(1:n-1)),2)\sim =r(n)$}
    {
     $r(Idx(1))= \sim r(Idx(1)));$    \tcp*{flip the least reliable bit}
    }
   $c=r;$
\caption{Proposed hard-decision decoder for PCC}\label{alg1}
\end{algorithm}
Note that Matlab's notation has been used in Algorithm \ref{alg1} so that $\sim =$ denotes `not equal to' and $\sim a$ denotes logical not of bit $a$.
\begin{thm}
The proposed hard-decision decoder has a diversity order of $d_{\min}=2$ for PCC.
\end{thm}
\begin{IEEEproof}
See Appendix A.
\end{IEEEproof}
While we have analyzed that the proposed decoder achieves full diversity for PCC by flipping one bit, we would like to explore whether we can flip more than one bit. Clearly, two bits cannot be flipped as that would not change the parity and hence the diversity order would be two or lower.
\section{Simulation results}
\label{sec:sim}

\begin{figure}[t]
	\begin{center}
		\includegraphics[width=3.5in,height=3.5in]{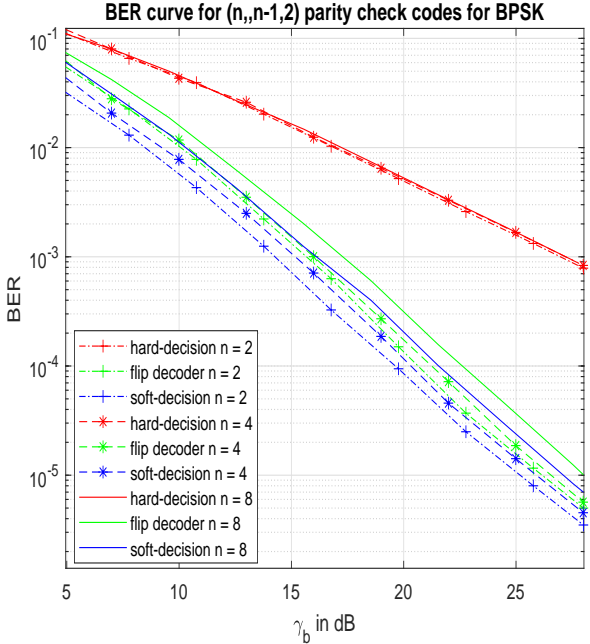}\\
		\caption{Simulation results for Parity check codes of length $n=$ 2, 4 and 8 over Rayleigh Fading Channels for hard-decision decoding, proposed Flip Decoding (FD) and soft-decision decoding.}
		\label{fig:BER}
	\end{center}
\end{figure}

Fig.~\ref{fig:BER} shows the BER vs. SNR per bit, $\gamma_b$, for hard-decision decoding, proposed flip decoding and soft-decision decoding over Rayleigh fading channels. The simulations were performed over code lengths of $n=2, 4 $ and 8 using BPSK modulation. Observe that the proposed FD decoder preserves diversity and that hard-decision decoding has a diversity order of one. Further, as $n$ increases from 2 to 8, the performance of FD and soft-decision decoding shifts by about 2 dB. Also, note that the performance of the proposed decoder is within 2 dB of soft-decision decoding. Simulations were also done using 4-QAM, 16-QAM and 64-QAM with similar conclusions and so have not been reported. Note that the complexity of brute-force ML soft-decision and hard-decision decoding is exponential in $n-1$. Also for PCC, as $t=\lfloor \frac{d_{\min}-1}{2}\rfloor=0$, the chase decoder \cite{4} and its variants reduce to hard-decision decoding. OSA and its variants \cite{25} have near-optimal performance and matches soft-decision decoding but  have at-least quadratic complexity in $n$ due to Gaussian elimination.  The proposed hard-decision decoder`s complexity involves finding the minimum of a vector that is of
linear complexity and then flipping the least reliable bit, which
is of constant complexity. So the overall complexity of the flip
decoder is linear in $n$. Also, note that the PCC rate, $1-1/n$, approaches one as $n$ increases. So this scheme is suitable for high data rate communication that requires a diversity order of two.
\section{Conclusions}
\label{sec:cncl}
For the first time, we have proposed a linear complexity hard-decision based decoding scheme that preserves diversity for any error correcting code; the parity check codes. The diversity performance of the decoding scheme was verified by analysis and simulations. Further enhancement of this work for all codes or codes with higher code distance will be an interesting topic of research. Especially in the context of ultra-reliable low latency communications (URLLC) and internet of things (IoT) \cite{26}, which require low complexity decoding with high performance gains.

\section*{Acknowledgments}
This publication is an outcome of the R$\&$D work under the Visvesvaraya Faculty research fellowship Ph.D. Scheme of Ministry of Electronics $\&$ Information Technology, Government of India; implemented by Digital India Corporation.

\setcounter{equation}{16}
\begin{table*}[h]
\begin{eqnarray}\label{eq12}
P_{FD}(c)&\leq&\frac{n^2(n-1)^2}{[\bar{\gamma}_c+(n-1)](\bar{\gamma}_c+n)\{\bar{\gamma}_c+1\}} \left\{1+\sum_{k=1}^{n-1} \frac{\binom{n-1}{k}\left(-1\right)^{k}\{\bar{\gamma}_c+1\}}{\bar{\gamma}_c+(k+1)}\right\}+ \sum_{m=2}^{n}\binom{n}{m}\left\{\frac{1}{1+\bar{\gamma}_c}\right\}^m\left\{\frac{\bar{\gamma}_c}{1+\bar{\gamma}_c}\right\}^{n-m}
\end{eqnarray}
\end{table*}
\setcounter{equation}{6}
\section*{Appendix A \\ Proof of Theorem 1}
For the proposed decoder,  an error occurs if  there  are more than one errors or  when there is one error but not in the bit with smallest $\gamma_i$. We denote the probability of one error but not in the bit with smallest $\gamma_i$ as $P_e(\mbox{error in 1 bit})$ and the probability of more than one errors as $P_e(\mbox{error in 2 or more bit})$. Accordingly, the probability of error is given by
\begin{equation}\label{eq4a}
  P_{FD}(c) = P_e(\mbox{error in 1 bit}) + P_e(\mbox{error in 2 or more bit}).
\end{equation}
The probability of getting two or more errors, given unordered $\gamma_i$'s is obtained as
\begin{equation}\label{eq4b}
 P_e(\mbox{error in 2 or more bit}) = \sum_{m=2}^{n}\binom{n}{m}\bar{p}^m(1-\bar{p})^{n-m},
\end{equation}
where $\bar{p}$ is the probability of error in a random bit corresponding to unordered $\gamma_i$'s, defined as
\begin{equation}\label{eq4c}
 \bar{p}=\mathbf{E}_{\gamma}\left[Prob\left\{n\ge {\gamma} | {\gamma} \right\}\right]=\mathbf{E}_{\gamma}\left[Q\left(\sqrt{2{\gamma}}\right)\right]
 \leq 
 \frac{1}{1+\bar{\gamma}_c},
\end{equation}
where the last inequality is obtained using the Chernoff bound \cite{Pro} and averaging over the exponential distribution of $\gamma$. Substituting \eqref{eq4c} in \eqref{eq4b} we obtain
\begin{align}\label{eq4d}
P_e(\mbox{error in 2 or more bit})= \sum_{m=2}^{n}\binom{n}{m}\left\{\frac{1}{1+\bar{\gamma}_c}\right\}^m\nonumber\\
\left\{\frac{\bar{\gamma}_c}{1+\bar{\gamma}_c}\right\}^{n-m}.
\end{align}
For finding $P_e(\mbox{error in 1 bit})$ the $\gamma_{i}$ 's are ordered, so that
$\gamma_{1} \leq \gamma_{2} \leq \cdots \leq \gamma_{n}$, define
\begin{equation}\label{eq4e}
 p_i=Prob\left\{n_i\ge {\gamma}_i | {\gamma}_i \right\}=Q\left(\sqrt{2{\gamma}_i}\right)\leq e^{-{{\gamma}_i}},
\end{equation}
where the last inequality is obtained using the Chernoff bound \cite{Pro}. Let $\bar{p}_i = \mathbf{E}_{\gamma_i}\left[p_i\right]$. This implies
$\bar{p}_{1} \geq \bar{p}_{2} \geq \cdots \geq \bar{p}_{n}$, {and} $\left(1-\bar{p}_{1}\right) \leq\left(1-\bar{p}_{2}\right) \leq \cdots \leq\left(1-\bar{p}_{n}\right)$.

Using the last inequality, we have
\begin{eqnarray}\label{eq5}
P_{e}\left(\mbox{ error in 1 bit }\right)&=&\sum_{i=2}^{n} \cdot \bar{p}_{i} \prod_{j=1 \atop j \neq i}^{n}\left(1-\bar{p}_{j}\right)
\nonumber\\
&\leq & \sum_{i=2}^{n} \bar{p}_{i}\left(1-\bar{p}_{n}\right)^{n-1} \nonumber\\ &\leq& (n-1) \bar{p}_{2}\left(1-\bar{p}_{n}\right)^{n-1},
\end{eqnarray}
where $\bar{p}_{2}$ and $\bar{p}_{n}$ are obtained by averaging over the ordered statistics of $\gamma_{i}$.
The pdf of the second order and the $n$-th order statistic corresponding to $\gamma_2$ and $\gamma_n$ are given by \cite{Ros}
\begin{eqnarray}\label{eq7}
f_{\gamma_{2}}(x)&=&n(n-1) \frac{1}{\bar{\gamma}_c}\left[e^{-\frac{x(n-1)}{\bar{\gamma}_c}}-e^{-\frac{xn}{\bar{\gamma}_c}}\right],\nonumber\\ f_{\gamma_{n}}(x)&=&n \frac{1}{\bar{\gamma}_c} e^{-\frac{x}{\bar{\gamma}_c}}\left[1-e^{-\frac{x}{\bar{\gamma}_c}}\right]^{(n-1)}
                  \nonumber\\&=&\sum_{k=0}^{n-1} \frac{n\binom{n-1}{k}(-1)^k}{\bar{\gamma}_c}e^{-\frac{x(k+1)}{\bar{\gamma}_c}}.
\end{eqnarray}
Evaluating the two expectations  in \eqref{eq5} separately, we have
\begin{eqnarray}
\bar{p}_{2} &\leq & \int_{0}^{\infty} n(n-1) \frac{1}{\bar{\gamma}_c}\left[e^{-\frac{x(n-1)}{\bar{\gamma}_c}}-e^{-\frac{xn}{\bar{\gamma}_c}}\right] e^{-x} d x\nonumber\\
&=&\int_{0}^{\infty} n(n-1) \frac{1}{\bar{\gamma}_c}\left[e^{-\frac{x(1+(n-1))}{\bar{\gamma}_c}}-e^{-\frac{x(1+n)}{\bar{\gamma}_c}}\right] d x\nonumber\\
&=& n(n-1)\left[\frac{1}{\bar{\gamma}_c+(n-1)}-\frac{1}{\bar{\gamma}_c+n}\right]\nonumber\\
&=&\frac{n(n-1)}{[\bar{\gamma}_c+(n-1)](\bar{\gamma}_c+n)}. \label{eq8}\\
\bar{p}_{n} &= & \int_{0}^{\infty} \sum_{k=0}^{n-1} \frac{n\binom{n-1}{k}(-1)^k  e^{-\frac{x(k+1)}{\bar{\gamma}_c}}}{\bar{\gamma}_c}  e^{-x } d x\nonumber\\
&=& \sum_{k=0}^{n-1} \frac{n\binom{n-1}{k}(-1)^k}{\bar{\gamma}_c+(k+1)}. \label{eq10}
\end{eqnarray}
Substituting \eqref{eq10}  and  \eqref{eq8}  in \eqref{eq5}, we have
\begin{eqnarray}\label{eq11}
P_{e}\left(\mbox{ error in 1 bit } \right)&\leq&\frac{n^2(n-1)^2}{[\bar{\gamma}_c+(n-1)](\bar{\gamma}_c+n)} \left\{\sum_{k=0}^{n-1}\right.\nonumber\\
&&\left. \left(-1\right)^{k} \frac{\binom{n-1}{k}}{\bar{\gamma}_c+(k+1)}\right\}.
\end{eqnarray}
Substituting from \eqref{eq11}, \eqref{eq4d} to \eqref{eq4a}, we have \eqref{eq12}.
\setcounter{equation}{17}
Using the definition of diversity order, $D$, \cite{Tse}, substituting from \eqref{eq12}  and applying the limits we have
\begin{IEEEeqnarray}{rCL}\label{eq13}
  D &=& \lim_{\bar{\gamma}_c \rightarrow \infty} \frac{-\log [P_{FD}(c)]}{\log[\bar{\gamma}_c]} \nonumber \\
   &=& \lim_{\bar{\gamma}_c \rightarrow \infty} \frac{2\log\left[1+\bar{\gamma}_c\right]+\mbox{higher order terms}}{\log[\bar{\gamma}_c]}=2.
\end{IEEEeqnarray}

\balance


\end{document}